\documentclass[superscriptaddress,groupedaddress,nobibnotes,twocolumn,amsmath,amssymb,prb]{revtex4-2}

\usepackage{placeins} 
\usepackage{tikz}
\usepackage{graphicx}
\usepackage{dcolumn}
\usepackage{bm}
\usepackage{hyperref}
\usepackage{comment}
\usepackage[normalem]{ulem}

\newcounter{subfigure}

\newcommand{\panref}[1]{Fig.~\ref{fig:Phonons}(\ref{#1})}
\begin{document}
\title{Singularly isostatic and geometrically unstable rigidity of metal-organic frameworks}%

\author{Christopher M. Owen}
\affiliation{%
Department of Physics, Applied Physics, and Astronomy, Binghamton University, Binghamton, NY 13902
}%

\author{Michael J. Lawler}
\affiliation{%
Department of Physics, Applied Physics, and Astronomy, Binghamton University, Binghamton, NY 13902
}%
\affiliation{%
Department of Physics, Cornell University, Ithaca, NY 14853
}%

\date{\today}

\begin{abstract}
Metal–organic frameworks (MOFs) combine high porosity with structural fragility, raising important questions about their mechanical stability. We develop a rigidity-based formalism in which spring networks parameterized by UFF4MOF are used to construct rigidity and dynamical matrices. Large-scale analysis of 5,682 MOFs from the CoRE MOF 2019 database shows that most frameworks are formally over-constrained yet cluster sharply near the isostatic threshold, revealing accidental geometric modes and placing many MOFs near mechanical instability. In the representative case of UiO-66, we show that auxiliary long-range constraints introduced by tuning the neighbor cutoff lift these modes into soft, flat, finite-frequency bands. The results show that rigidity-matrix analysis can rapidly identify MOFs likely to remain mechanically stable. This near-criticality mirrors behavior known from topological mechanics and points to a deeper design principle in porous crystals.
\end{abstract}

\maketitle



\section{Introduction}
\begin{figure}[t]
\centering
\includegraphics[width=\linewidth]{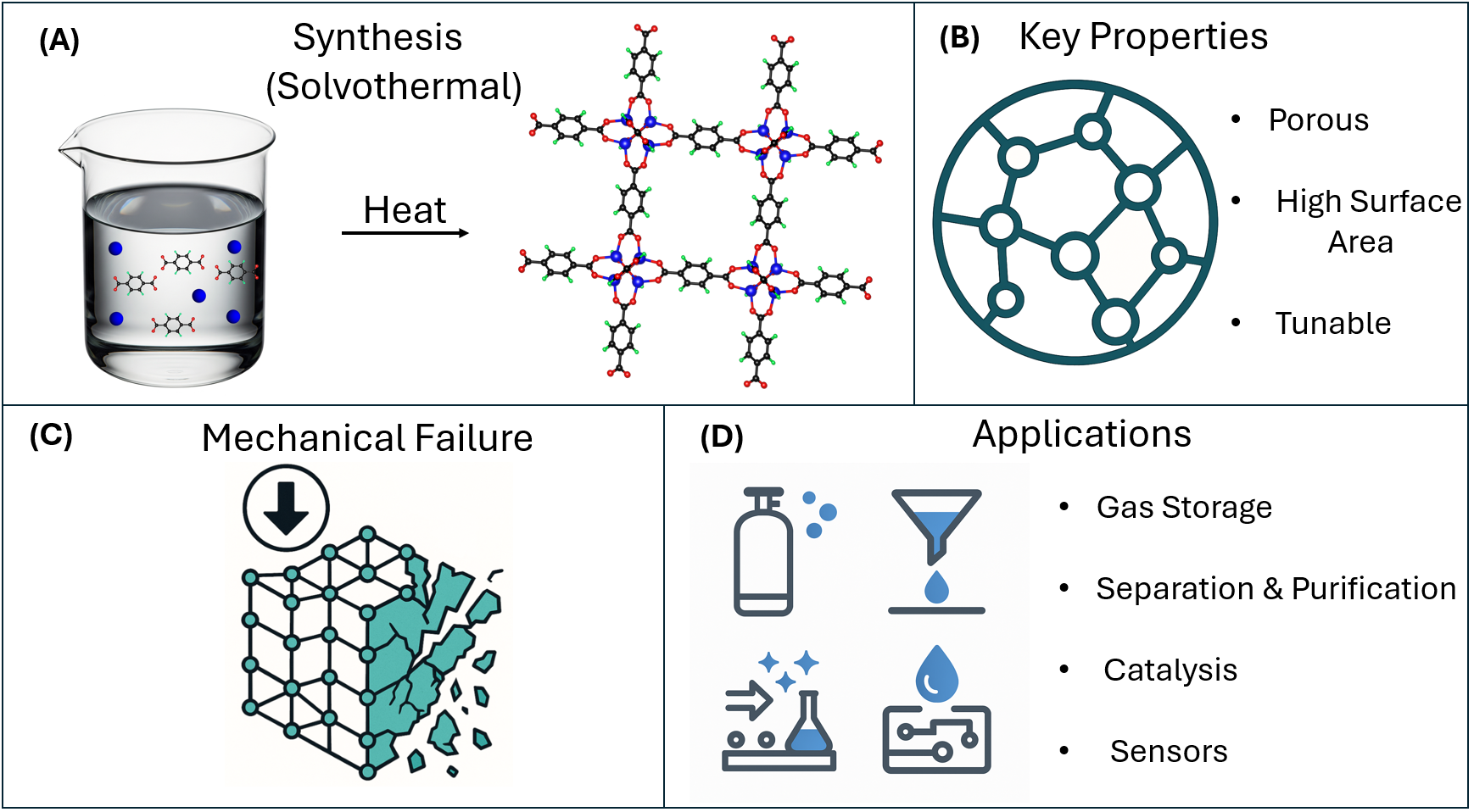}
\caption{\textbf{A}: MOFs are typically obtained via solvothermal methods~\cite{solvothermal}, where metal ions or clusters and multidentate organic linkers self-assemble into extended crystalline networks.
\textbf{B}: The resulting frameworks exhibit permanent porosity, high internal surface areas, and tunable structures and functionalities.
\textbf{C}: Many MOFs are mechanically fragile and can degrade or collapse under external stress, highlighting the importance of identifying mechanically robust frameworks.
\textbf{D}: MOFs have been investigated for gas storage, separations, heterogeneous catalysis, and chemical sensing.}
\label{fig:What is a mof}
\end{figure}

Metal--organic frameworks (MOFs) combine crystalline order with exceptional porosity, creating vast internal surface areas and highly tunable architectures that are promising for applications ranging from gas storage to catalysis (Fig.~\ref{fig:What is a mof})~\cite{Furukawa2010,Zhou-mofs}. These same attributes often place MOFs near the edge of mechanical~\cite{MOF-prop} stability: frameworks collapse during activation, deform under modest pressure, or exhibit soft, collective motions that couple strongly to guest molecules~\cite{ZEGGAI2025100864}. Predicting which structures are robust and which are mechanically fragile remains a central challenge~\cite{MOF-prop2,Moosavi2018ChemicalCaryatids} for design and screening. Early efforts to describe MOF mechanics include the hinge--truss model of Sarkisov~\cite{Sarkisov2014}, which showed that framework flexibility can often be anticipated from network topology. Marmier and Evans~\cite{MOFflex} subsequently carried out a more mathematical, group-theoretical analysis of MOF frameworks, demonstrating that simple scalar Maxwell--Calladine counts are unreliable for MOFs. Since self stress can obscure underlying mechanisms a more robust method is needed thus motivating a matrix-based approach. 

First principles and classical approaches each address part of this problem. Density functional theory (DFT) captures bonding and elastic responses with high accuracy but is limited to small cells~\cite{Maurer2019}. Classical force fields and machine-learned interatomic potentials (MLIPs)~\cite{Moghadam2019,Cleeton2025DeepDreamMOF,MOF-ML} extend accessible scales and enable dynamics for larger systems, but remain too computationally expensive for high-throughput screening and often yield descriptive metrics without revealing the structural degrees of freedom that govern rigidity.

An alternative to first-principle or machine-learning approaches is to construct a simplified mechanical model that captures the the essence of the material. One established approach in this direction is the Rigid Unit Model (RUM), which identifies low energy vibrational modes by treating polyhedra in crystalline solids as perfectly stiff bodies connected by flexible hinges ~\cite{RUM}. By searching for geometric distortions that leave these rigid units undeformed, the RUM model successfully predicts low-energy vibrational modes and flexibility windows based solely on connectivity. For disordered solids, the analytical theory of nonaffine deformation ~\cite{Affine} is insightful. This formalism explicitly links the vibrational spectrum to mechanical stability by showing how nonaffine displacements reduce the shear modulus, thus driving the material toward a loss of rigidity as the coordination number decreases. The affine stiffness is calculated by averaging the geometric orientations of the atomic bonds. Both of these approaches utilize geometry to gain insight.

We take a complementary route based on rigidity theory, also known as topological constraint theory (TCT)~\cite{Thorpe1983,Jacobs1995}. In this formalism, a structure's mechanical response is governed by the balance between internal degrees of freedom and the independent constraints imposed by its bonding network. The key objects are zero modes which are motions that cost no energy and states of self stress, which are patterns of internal tension and compression that satisfy force balance. The Maxwell--Calladine~\cite{Maxwell1864,Calladine1978} relation, 
\begin{equation}\label{eq:MC} 
    \nu \equiv N_0 - N_{\mathrm{ss}} = dN_{\mathrm{s}} - N_{\mathrm{c}}, 
\end{equation}
links these quantities to network connectivity.
In molecular terms, \(dN_{\mathrm{s}}\) is the total atomic degrees of freedom and \(N_{\mathrm{c}}\) is the number of linear bond stretching and angular bond bending constraints; together they fix the number of zero modes (\(N_0\)) and states of self stress (\(N_{\mathrm{ss}}\)). 

This constraint-based approach offers three advantages that are hard to obtain together with other methods. It clarifies which motions are allowed by the network topology, scales to system sizes suitable for high-throughput screening, and helps distinguish mechanically robust frameworks from those prone to soft modes or collapse.

In this work, we operationalize this program for MOFs by building constraint-based Hamiltonians with linear and angular springs, assembling rigidity matrices for periodic crystals, computing their Maxwell--Calladine indices, and phonon dispersions. The resulting modes are then characterized using the Inverse Participation Ratio (IPR) to distinguish between localized and collective motions, which is crucial for identifying the origins of framework flexibility. Spring constants are assigned from UFF4MOF~\cite{Rappe1992,Addicoat2014,UFF-Ext} parameters, and all analyses are performed in reciprocal space to resolve wave vector-dependent zero modes. Applied to 5,682 frameworks in the CoRE 2019~\cite{Chung2024CoREMOF} dataset, this approach reveals that most MOFs occupy a mechanically marginal regime while only a smaller subset behaves as genuinely rigid frameworks. In this way, the rigidity-matrix analysis complements first-principles benchmarks: DFT and MLIPs provide accuracy for selected targets, while the present approach provides scalable guidance and interpretable relations across thousands of candidates.

\begin{figure} [h!]
    \centering
    \includegraphics[width=\linewidth]{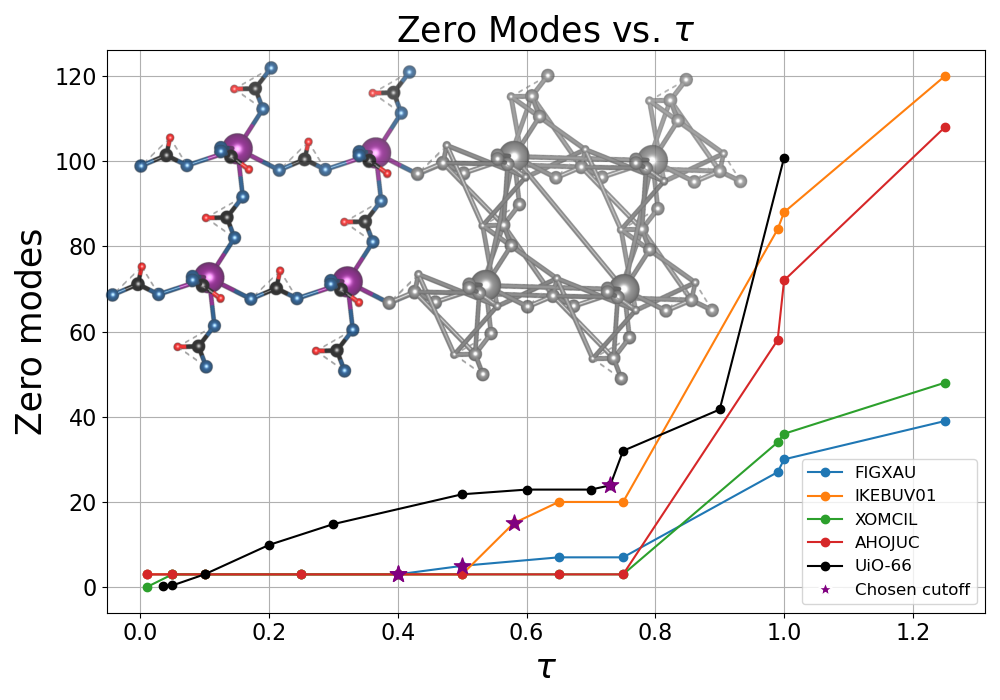}
    \caption{Zero-mode count $N_0$ as a function of the constraint cutoff parameter $\tau$, which controls which neighbors are included in the rigidity model. In our convention, larger $\tau$ reduces the coordination weight used by \texttt{CrystalNN}\cite{crystalNN}, removing constraints and increasing $N_0$. Each colored curve corresponds to a different MOF, and purple stars ($\ast$) mark the plateau $\tau$ values selected automatically by CrystalNN. For clarity, the UiO-66 curve is scaled by a factor of 10. The inset shows the FIGXAU framework: Color indicates the baseline bonding network, while the gray version illustrates the auxiliary bonds identified when $\tau$ is reduced.}
    \label{fig:Tau}
\end{figure}

\section{Methodology}

\subsection{Constraint-Based Model and Parameterization}

We model each MOF as a periodic network of atoms connected by harmonic springs that penalize bond stretching and bond bending. The total Hamiltonian is
\begin{equation}\label{eq:H}
    H = \sum_{i=1}^N \frac{p_i^2}{2m_i}
      + \tfrac{1}{2}\sum_{\mu} K_\mu \,(e_\mu)^2 
\end{equation}
where $\mu$ indexes all independent constraints in the network, including both bond-stretching terms $(i,j)\in\mathcal{L}$ and bond-bending terms $(i,j,k)\in\mathcal{A}$. The associated extensions are $e_{ij} = \hat{\mathbf{e}}_{ij}\cdot(\mathbf{u}_i-\mathbf{u}_j)$ and $e_{ijk} = \theta_{ijk} - \theta^0_{ijk}$. Stacking all extensions into a vector $\mathbf{e}$ and displacements into $\mathbf{u}$ defines the rigidity matrix $R$ through $\mathbf{e}=R\mathbf{u}$. In our implementation, the same $R$ used for Maxwell–Calladine counting also enters the energetic dynamical matrix once local stiffness and prestress are specified, consistent with the generalized rigidity analysis of Rocks and Mehta~\cite{rocks2024}. The potential energy and dynamical matrix then follow as:

\begin{equation}
    V = \tfrac{1}{2}\,\mathbf{u}^T (R^T K R)\,\mathbf{u} \\
\end{equation}
\begin{equation}
    D= M^{-1/2}\,R^T K\,R\,M^{-1/2}
\end{equation}

Here $M$ is the diagonal mass matrix constructed from the atomic masses $m_i$, and $K$ are the bond and angle spring constants obtained from UFF4MOF parameters, with atom types assigned according to local coordination environments. Explicitly,

\begin{align}
    K_{ij} &= 664.12 \,\frac{Z_i Z_j}{r_{ij}^3} \\
    K_{ijk} &= \beta \;\frac{Z_i Z_k}{r_{ik}^5}\,
    \Big[\,3\,r_{ij}r_{jk}(1-\cos^2\theta_0)-r_{ik}^2\cos\theta_0\Big],
\end{align}

where $\beta = 664.12/(r_{ij}r_{jk})$ and $Z_\ell$ are the UFF effective charges. We focus on these terms as they provide the dominant constraints governing framework rigidity. The UFF equilibrium distances are used as rest lengths; across all linear bonds they exceed the geometric bond lengths by 5–7\% on average (see Appendix~\ref{Appendix:Data}).

The bonding network itself is determined using the \texttt{CrystalNN}\cite{crystalNN} algorithm implemented in \texttt{pymatgen}\cite{pymatgen}. A tunable cutoff parameter $\tau$ controls which coordination shells are included, as shown in {Fig.~\ref{fig:Tau}}. In the main dataset analysis we adopt the plateau $\tau$ identified by \texttt{CrystalNN}, which provides a chemically sensible minimal network, but in Fig.~\ref{fig:Ui-kicked} we explore the effect of reducing $\tau$ for UiO-66, showing that modest additions of longer-range constraints systematically lift many of the baseline zero modes into soft, finite-frequency vibrations.

\subsection{Phonon Dispersions and Localization Analysis}

For periodic crystals, the model is solved in reciprocal space. Atomic displacements are expanded as plane waves (see Appendix~\ref{Appendix:FT}),
\[
  \mathbf{u}_{m\alpha} = \frac{1}{\sqrt{N}}\sum_{\mathbf{k}}
  \mathbf{u}_\alpha(\mathbf{k})\,e^{i\mathbf{k}\cdot\mathbf{R}_m},
\]
which yields the $\mathbf{k}$-dependent dynamical matrix $D(\mathbf{k})$. Diagonalizing $D(\mathbf{k})$ gives the phonon spectrum $\omega^2(\mathbf{k})$, which we compute along the Setyawan--Curtarolo~\cite{Setyawan_2010} high-symmetry paths in the Brillouin zone.

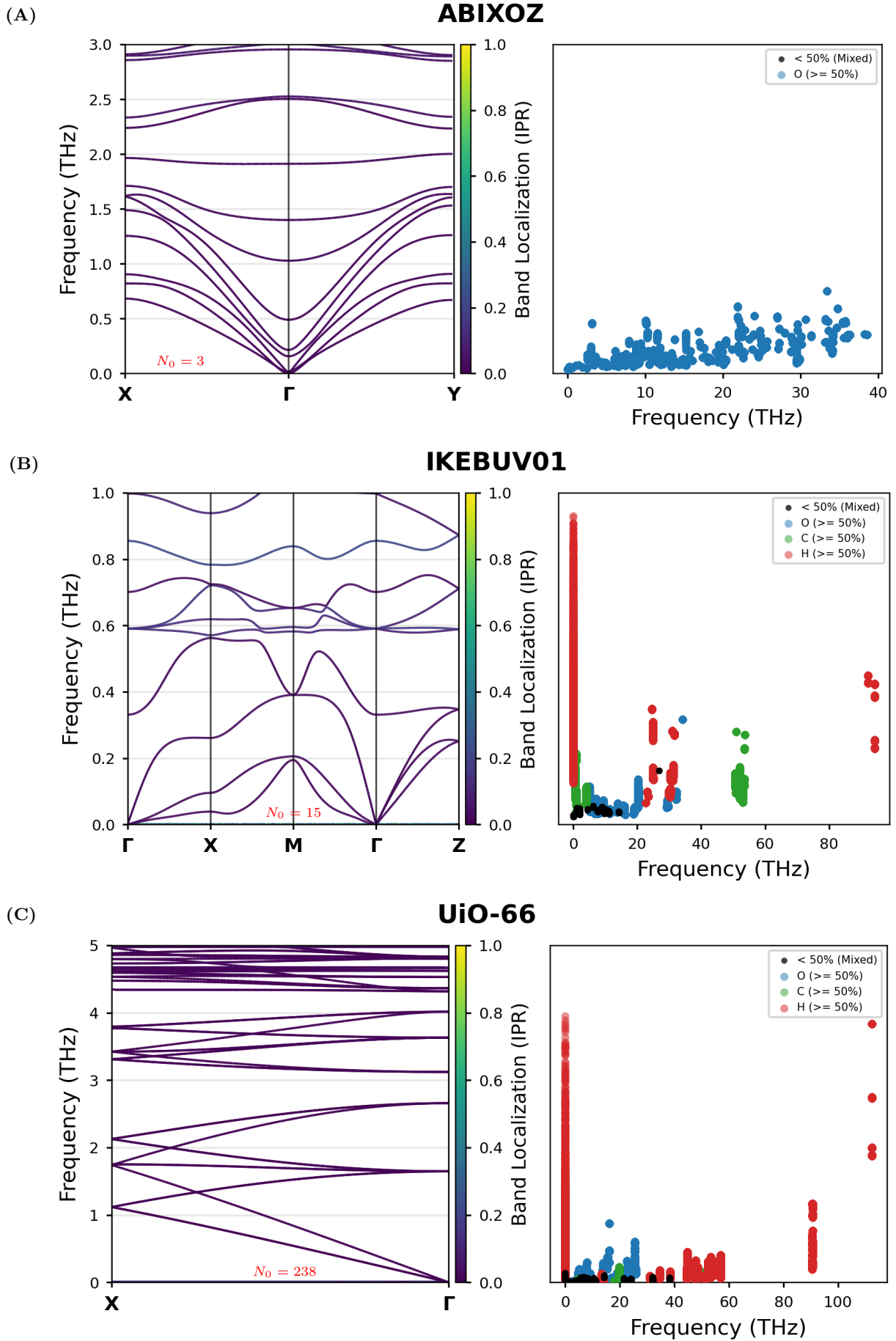
\begin{figure*}[p]
\centering
\setcounter{subfigure}{0}
\refstepcounter{subfigure}\label{fig:ABIXOZ}
\begin{tikzpicture}
\node[inner sep=0] (img) {\includegraphics[width=0.85\textwidth,trim=4 6 4 2,clip]{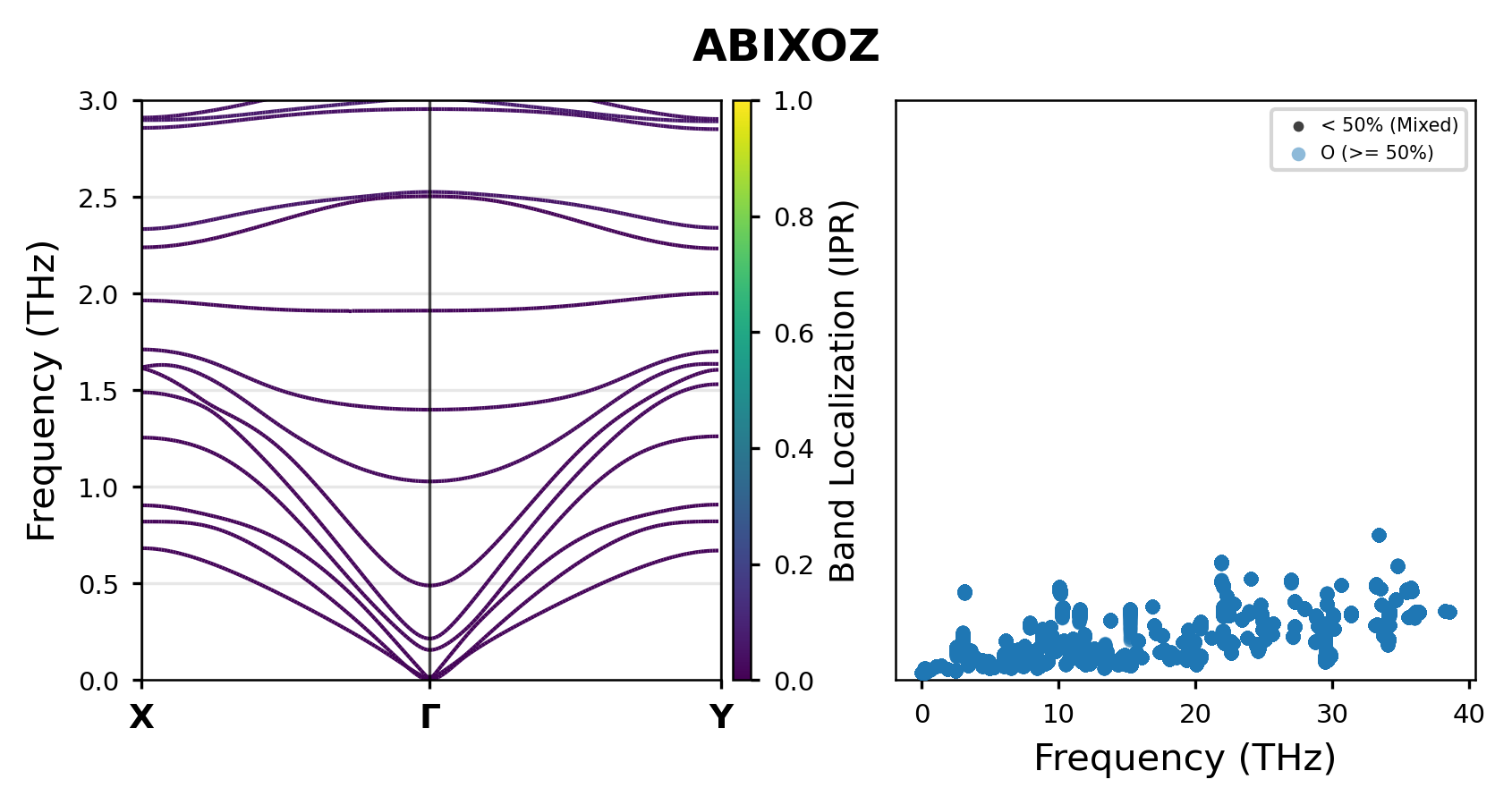}};
\node[anchor=north east, fill=white, inner sep=2pt]
at ([xshift=-6pt,yshift=-6pt]img.north west) {\bfseries (A)};
\node[text=red, font=\scriptsize, anchor=south west]
at ([xshift=50pt,yshift=30pt]img.south west) {$N_0 = 3$};
\end{tikzpicture}
\vspace{0.4em}

\refstepcounter{subfigure}\label{fig:IKEBUV01}
\begin{tikzpicture}
\node[inner sep=0] (img) {\includegraphics[width=0.85\textwidth,trim=4 6 4 2,clip]{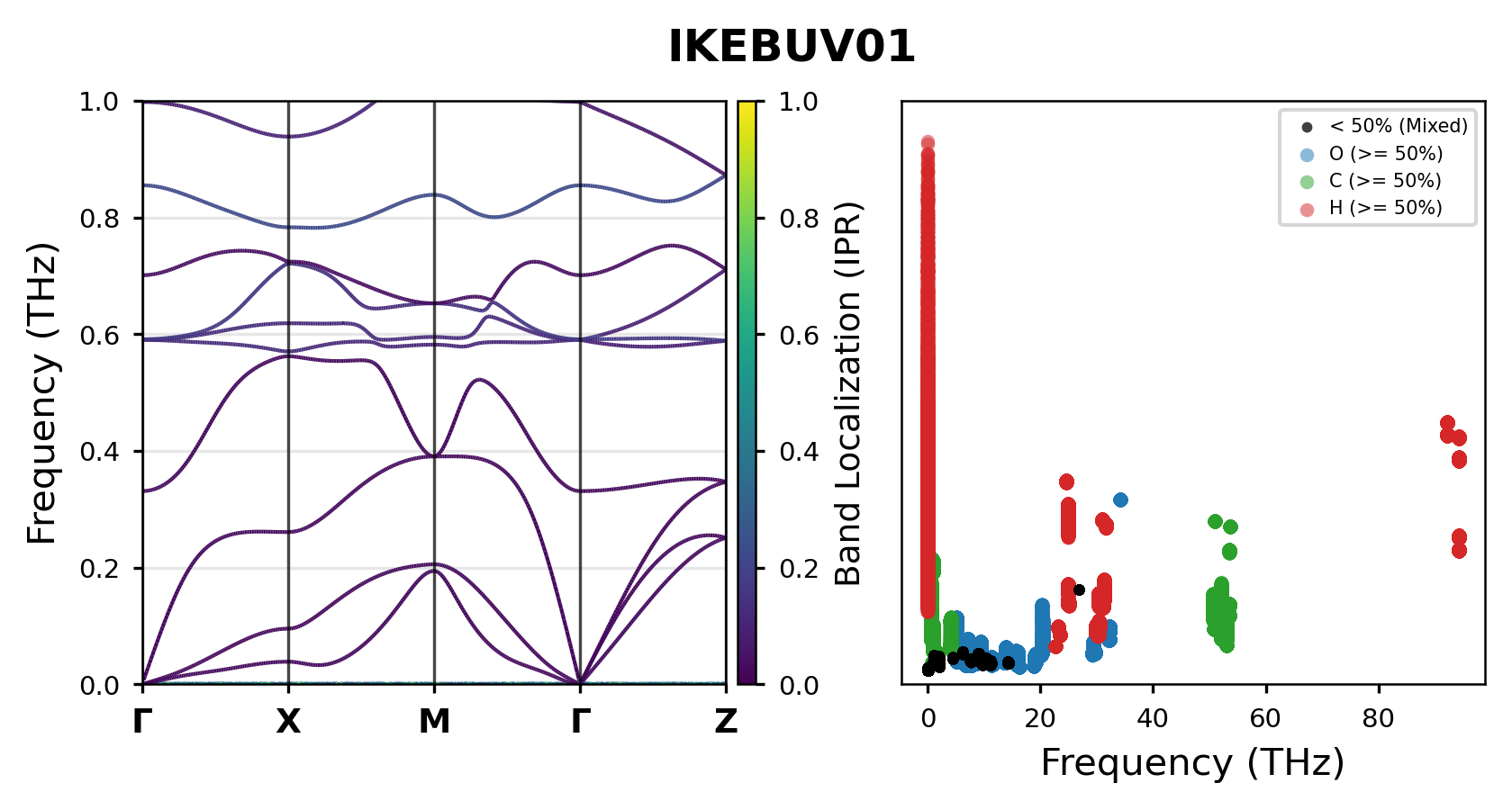}};
\node[anchor=north east, fill=white, inner sep=2pt] at ([xshift=-6pt,yshift=-6pt]img.north west) {\bfseries (B)};

\node[text=red, font=\scriptsize, anchor=south west]
at ([xshift=105pt,yshift=30pt]img.south west) {$N_0 = 15$};
    \end{tikzpicture}

    \vspace{0.4em}
    \refstepcounter{subfigure}\label{fig:UiO66}
    \begin{tikzpicture}
        \node[inner sep=0] (img) {\includegraphics[width=0.85\textwidth,trim=4 6 4 2,clip]{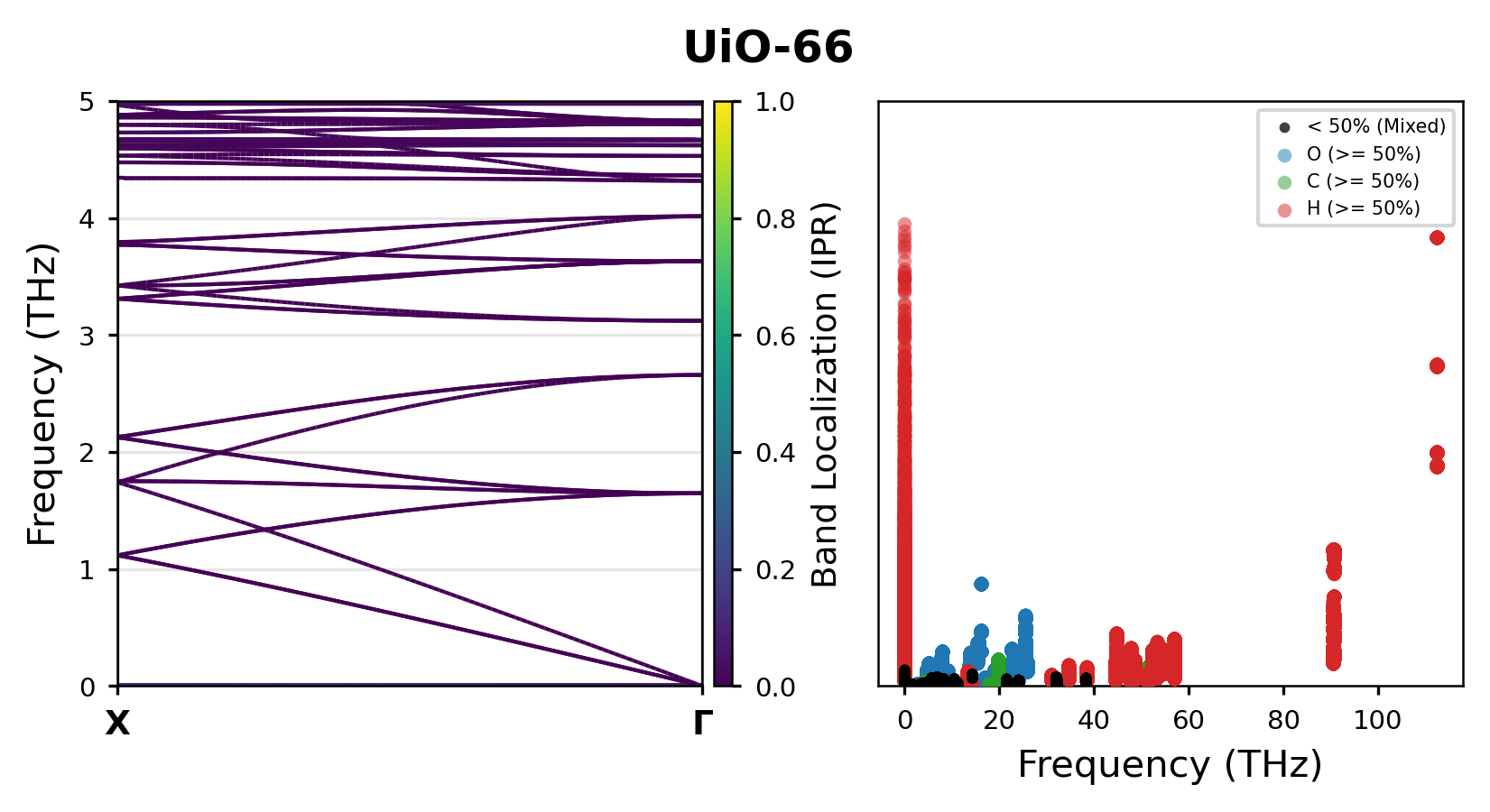}};
        \node[anchor=north east, fill=white, inner sep=2pt]
              at ([xshift=-6pt,yshift=-6pt]img.north west) {\bfseries (C)};
        \node[text=red, font=\scriptsize, anchor=south west]
at ([xshift=100pt,yshift=30pt]img.south west) {$N_0 = 238$};
    \end{tikzpicture}
    \vspace{-0.4em}
    \caption{Phonon dispersions and inverse participation ratio (IPR) maps for representative materials.
    \textbf{(A)} ABIXOZ ($\nu/N_s \!=\! -1.89$) stable and over-constrained;
    \textbf{(B)} IKEBUV01 ($\nu/N_s \!=\! +0.075$) near-isostatic with delocalized soft modes;
    \textbf{(C)} UiO-66 ($\nu/N_s \!=\! -1.14$) moderately over-constrained with mixed localization behavior.}
    \label{fig:Phonons}
\end{figure*}

\begin{figure*}[t]
    \centering
    \begin{tikzpicture}
       
        \node[anchor=south west,inner sep=0] (img) 
            {\includegraphics[width=\linewidth]{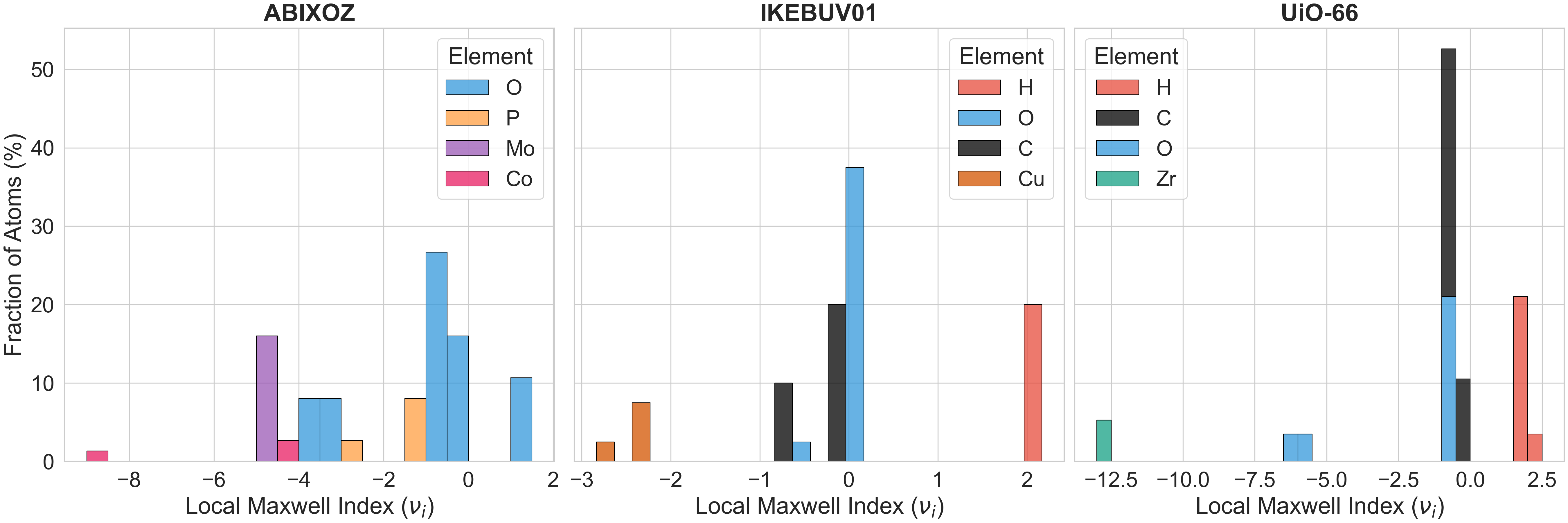}};
        
        \begin{scope}[x={(img.south east)},y={(img.north west)}]
            
            \node[anchor=north west,font=\bfseries] at (0.03,1.1) {(A)};
            \node[anchor=north,font=\bfseries]      at (0.38,1.1) {(B)};
            \node[anchor=north east,font=\bfseries] at (0.73,1.1) {(C)};
        \end{scope}
    \end{tikzpicture}
    \caption{Normalized distributions of the local Maxwell index $\nu_i$ for the three representative MOFs in Fig.~\ref{fig:Phonons}. Positive $\nu_i$ indicates locally under-constrained environments, $\nu_i\approx 0$ near-isostatic environments, and negative $\nu_i$ locally over-constrained environments. ABIXOZ (\textbf{A}) exhibits a comparatively uniform, over-constrained distribution, whereas IKEBUV01 (\textbf{B}) and UiO-66 (\textbf{C}) show strong heterogeneity with under-constrained hydrogen-rich environments coexisting with a near-isostatic framework backbone (dominated by heavier atoms) UiO-66 additionally exhibits a strongly over-constrained metal-node population at large negative $\nu_i$.}
    \label{fig:LocalR}
\end{figure*}

To distinguish between localized and delocalized vibrational modes, we calculate the inverse participation ratio (IPR). For a phonon mode at wavevector $\mathbf{k}$ and band index $j$, the participation of atom $i$ is $P_i(\mathbf{k}, j) = |\mathbf{u}_i(\mathbf{k}, j)|^2$. The IPR is then
\begin{equation}
    \mathrm{IPR}(\mathbf{k}, j)
    = \frac{\sum_{i=1}^{N} [P_i(\mathbf{k}, j)]^2}
           {\left( \sum_{i=1}^{N} P_i(\mathbf{k}, j) \right)^2}
\end{equation}
Low IPR values (approaching $1/N$) indicate delocalized modes, whereas high values (approaching 1) signal localization. In degenerate subspaces we choose a random orthonormal basis, which provides a representative sampling of possible localization patterns.

\subsection{Dataset and Implementation Details}

We apply this method to 5,682 structures from the CoRE MOF 2019 database~\cite{Chung2024CoREMOF}. Primitive cells are used whenever available to reduce redundancy. CIF~\cite{CIF_note} parsing, neighbor detection, constraint assembly, and phonon analysis were automated to enable large-scale screening. To validate our method, we benchmarked the rigidity code against crystalline silicon, shown in Appendix~\ref{Appendix:Phonons}, and against UiO-66, whose detailed analysis is presented in the main text. Full implementation details and all input files are provided in Appendix~\ref{Appendix:Data} and in the publicly available repository referenced therein.

\begin{figure*}[htbp!]
\centering
\begin{minipage}[t]{0.24\textwidth}
\centering
\begin{tikzpicture}
\node[inner sep=0] (img)
{\includegraphics[width=\linewidth,trim=4 6 4 2,clip]{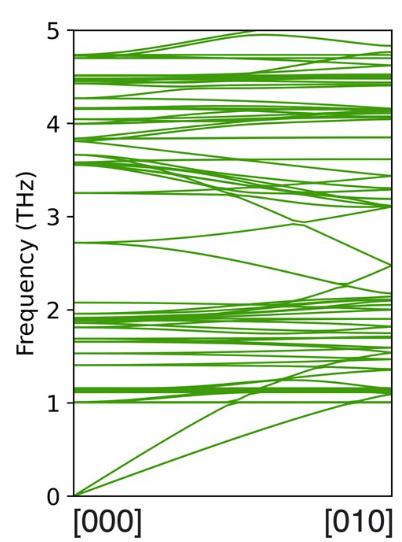}};
\node[anchor=north west, fill=white, inner sep=2pt]
at ([xshift=-6pt,yshift=11pt]img.north west) {\bfseries (A)};
\end{tikzpicture}
\end{minipage}
\hfill
\begin{minipage}[t]{0.24\textwidth}
\centering
\begin{tikzpicture}
\node[inner sep=0] (img)
{\includegraphics[width=\linewidth,trim=4 6 4 2,clip]{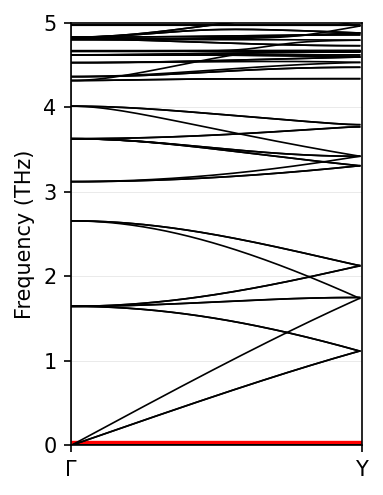}};
\node[anchor=north west, fill=white, inner sep=2pt]
at ([xshift=-6pt,yshift=16pt]img.north west) {\bfseries (B)};
\node[text=red, font=\scriptsize, anchor=south west]
at ([xshift=75pt,yshift=20pt]img.south west) {$N_0 = 238$};
\end{tikzpicture}
\end{minipage}
\hfill
\begin{minipage}[t]{0.24\textwidth}
\centering
\begin{tikzpicture}
\node[inner sep=0] (img)
{\includegraphics[width=\linewidth,trim=4 6 4 2,clip]{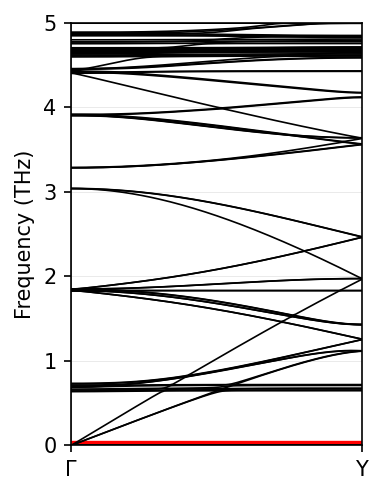}};
\node[anchor=north west, fill=white, inner sep=2pt]
at ([xshift=-6pt,yshift=16pt]img.north west) {\bfseries (C)};
\node[text=red, font=\scriptsize, anchor=south west]
at ([xshift=75pt,yshift=16pt]img.south west) {$N_0 = 215$};
\end{tikzpicture}
\end{minipage}
\hfill
\begin{minipage}[t]{0.24\textwidth}
\centering
\begin{tikzpicture}
\node[inner sep=0] (img)
{\includegraphics[width=\linewidth,trim=4 6 4 2,clip]{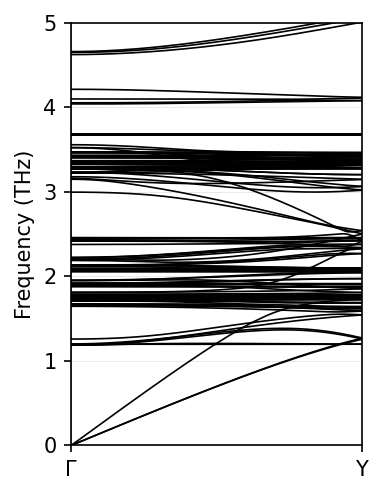}}; 
\node[anchor=north west, fill=white, inner sep=2pt]
at ([xshift=-6pt,yshift=16pt]img.north west) {\bfseries (D)};
\node[text=red, font=\scriptsize, anchor=south west]
at ([xshift=75pt,yshift=16pt]img.south west) {$N_0 = 0$};
\end{tikzpicture}
\end{minipage}

\caption{
(\textbf{A}) Benchmark phonon dispersion calculated via GULP, reproduced from Ref.~\cite{UiO-66}.
(\textbf{B}) Corresponding result from our rigidity-based model ($N_0 = 238$) using the default $\tau$ setting.
(\textbf{C}) Decreasing the cutoff to $\tau = 0.5$ introduces auxiliary constraints, reducing the zero mode count to $N_0 = 215$.
(\textbf{D}) Further tuning to $\tau = 0.035$ eliminates all zero modes ($N_0 = 0$). The progression from (\textbf{B}) to (\textbf{D}) demonstrates how assigning finite stiffness to long-range auxiliary bonds lifts zero modes into soft, collective bands, progressively filling the low-frequency spectrum relative to the benchmark in (\textbf{A}).
}
\label{fig:Ui-kicked}
\end{figure*} 

\begin{figure}[t]
    \centering

    \begin{tikzpicture}
        \node[inner sep=0] (img) {\includegraphics[width=\linewidth,trim=4 6 4 2,clip]{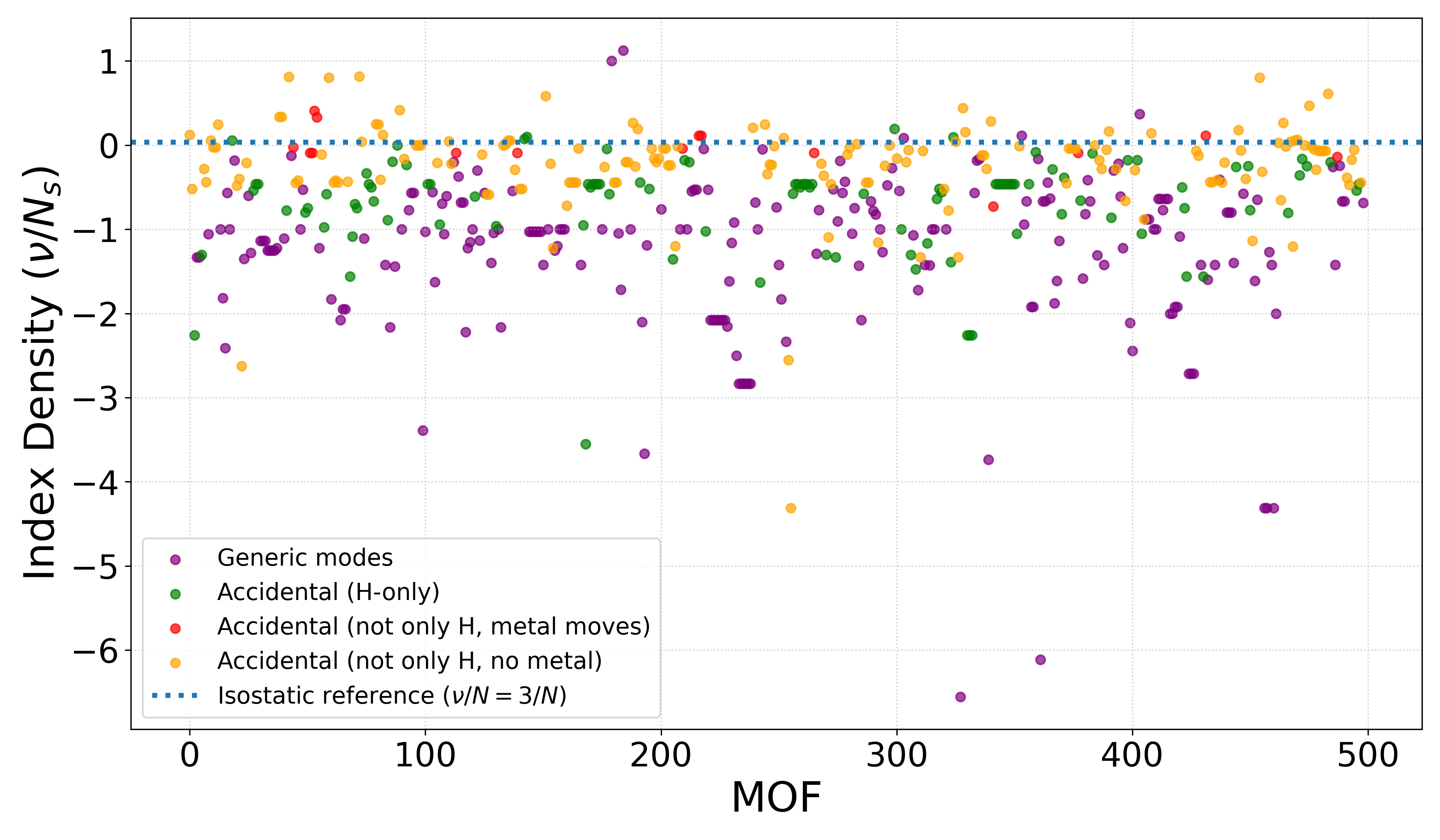}};
        \node[anchor=north west, fill=white, inner sep=2pt]
              at ([xshift=-8pt,yshift=-0pt]img.north west) {\bfseries (A)};

    \end{tikzpicture}

    \vspace{0.4em}

    \begin{tikzpicture}
        \node[inner sep=0] (img) {\includegraphics[width=\linewidth,trim=4 6 4 2,clip]{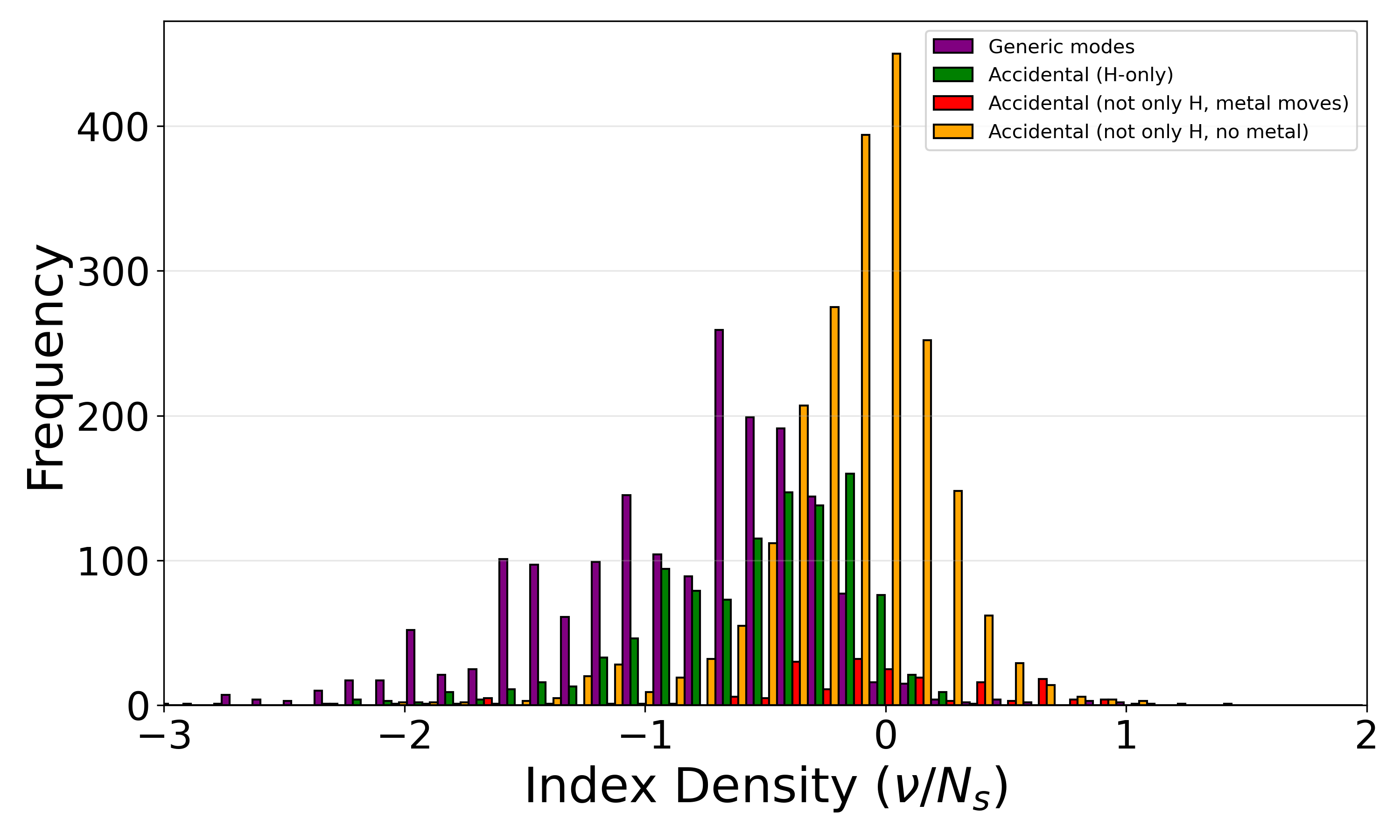}};
        \node[anchor=north west, fill=white, inner sep=2pt]
              at ([xshift=-8pt,yshift=-0pt]img.north west) {\bfseries (B)};
    \end{tikzpicture}

    \vspace{0.3em}

    \caption{Rigidity statistics across the CoRE MOF dataset.
    (\textbf{A}) Scatter plot of $\nu$ for 500 representative MOFs, with the blue dashed line marking the isostatic threshold ($\nu = 3$).
    (\textbf{B}) Histogram of the Maxwell–Calladine index density ($\nu/N_s$) for 5,182 MOFs at $\Gamma$. 
    The majority of frameworks are formally over-constrained yet cluster near the isostatic point.
    This concentration is dominated by accidental modes, which place many MOFs on the verge of mechanical instability.}
    \label{fig:density}
\end{figure}

\section{Results}

Our large-scale analysis of 5,682 structures from the CoRE MOF 2019 database shows that most frameworks are formally over-constrained at $\Gamma$. Figure~\ref{fig:density} summarizes this dataset-level behavior in two complementary views. The histogram shows that the Maxwell--Calladine index density $\nu/N_s$ is typically negative, indicating an excess of constraints relative to degrees of freedom. However, rather than being broadly distributed, the population concentrates strongly near the isostatic reference $\nu/N_s = 3/N$, which lies close to zero for large unit cells. The scatter plot shows that this near-isostatic concentration is widespread across the dataset rather than being driven by a small number of outliers.

Partitioning the statistics by zero-mode character shows that the dominant contribution to this near-isostatic peak comes from \emph{accidental} modes. The largest share arises from accidental modes that are not confined to hydrogen yet also do not substantially involve metal-node motion. Hydrogen-only accidental modes form a secondary contribution, while accidental modes that significantly move metal atoms are comparatively rare.

This sharp concentration indicates a universal mechanical trend: the self-assembly of porous frameworks drives many MOFs toward the brink of mechanical instability. In our dataset, this marginal behavior is dominated by \textit{accidental geometric modes}, which arise when the generic-geometry assumption behind Maxwell--Calladine counting is violated. The local geometry of metal clusters and linkers can create special alignments that render some constraints redundant, producing additional zero modes despite an overall excess of constraints. In contrast, frameworks classified as generic are shifted toward more negative $\nu/N_s$ values, consistent with conventional stability from independent constraints.
To systematically characterize these behaviors, we classify frameworks into three regimes based on the relationship between $\nu$ and the vibrational spectrum:
\begin{itemize}
    \item \textbf{Generic Rigidity ($\nu < 3$, $N_0 = 3$):} All constraints are independent. Topology dictates stability, and only global rigid-body zero modes occur.
    \item \textbf{Geometrically Unstable ($\nu < 3$, $N_0 > 3$):} Symmetry creates constraint redundancy, producing accidental nontrivial mechanisms despite being formally over-constrained.

    \item \textbf{Singular Isostatic (accidental, $\nu = 3$):}
An accidental isostatic case in which internal zero modes and states of self stress coexist, with $N_0 > 3$, $N_{\mathrm{ss}} > 0$, and $N_0 - N_{\mathrm{ss}} = 3$.

\end{itemize}

Structures with $\nu>3$ are classified as under-constrained and are stabilized only by weak long-range couplings. In practical terms, generically rigid frameworks are robust to small distortions, whereas geometrically unstable and singular isostatic frameworks can be pushed into rigid behavior by modest geometric changes. We illustrate these regimes using the representative MOFs ABIXOZ, IKEBUV01, and UiO-66 in Figs.~\ref{fig:Phonons} and \ref{fig:LocalR}: Fig.~\ref{fig:Phonons} shows phonon dispersions together with inverse participation ratio (IPR) maps labeled by the dominant atomic species in each eigenvector, while Fig.~\ref{fig:LocalR} resolves the corresponding atomic-scale constraint balance via the local Maxwell index distributions $\nu_i$. This local index is defined for each atom \textit{i} as:
\begin{equation}
    \nu_i = 3 - r_i
\end{equation}
where $r_i$ is the local constraint density. By partitioning the global constraints $\mu$ defined in Eq.~\eqref{eq:H}, we assign:
\begin{equation}
    r_i = \frac{1}{2} n_{i, \mathcal{L}} + \frac{1}{3} n_{i, \mathcal{A}}
\end{equation}
where $n_{i, \mathcal{L}}$ is the number of bonds in $\mathcal{L}$ connected to atom $i$, and $n_{i, \mathcal{A}}$ is the number of bond angles in $\mathcal{A}$ in which the atom participates. This partitioning ensures that the local indices sum to the global index, $\sum \nu_i = \nu$.

ABIXOZ represents the \emph{generic rigidity} regime. It is strongly over-constrained ($\nu/N_s=-1.89$) and contains no non-translation zero modes at $\Gamma$. Consistent with this, its dispersion in \panref{fig:ABIXOZ} is smooth and well ordered, and its IPR profile is predominantly delocalized with a stable dominant-species signature. The corresponding local-index distribution is comparatively uniform, consistent with a framework whose constraints are largely independent and whose rigidity is well predicted by global counting.

IKEBUV01 exemplifies the \emph{singular isostatic} regime. At the mechanical critical point ($\nu=3$), its stability is symmetry-enforced rather than generic: it contains twelve nontrivial zero modes at $\Gamma$ that are exactly canceled by twelve states of self stress. In \panref{fig:IKEBUV01}, this marginality appears as low-frequency asymmetries and nearly flat branches. The IPR map shows that the soft and zero-like motion at $\Gamma$ is overwhelmingly dominated by hydrogen, indicating that the accidental zero-mode subspace is largely confined to linker-associated, peripheral degrees of freedom rather than involving the load-bearing backbone. The local rigidity distribution in Fig.~\ref{fig:LocalR} mirrors this separation: hydrogen is concentrated at positive $\nu_i$, while the heavier framework atoms cluster near $\nu_i\approx0$, consistent with a near-isostatic backbone coexisting with locally under-constrained hydrogen-rich environments. Despite being mechanically marginal by symmetry, this composition suggests that many of the zero modes are likely mechanically benign internal flexibilities.

UiO-66 lies in the \emph{geometrically unstable} regime. Despite being globally over-constrained ($\nu/N_s=-1.14$), it exhibits 238 zero modes at $\Gamma$, indicating accidental geometric modes produced by constraint dependencies. In this zero-mode subspace, motion is dominated by hydrogen (about $83\%$ participation), with a substantial carbon component (about $16\%$) and only minor oxygen participation (about $1\%$), indicating that the soft degrees of freedom primarily reside on ligand-associated hydrogen and carbon environments rather than on the metal--oxygen backbone. In \panref{fig:UiO66}, these appear as bands that flatten near $\Gamma$, signaling compliant internal degrees of freedom within an otherwise over-constrained network; the associated IPR patterns show strong hydrogen participation at low frequency together with mixed-character modes at higher frequency. Figure~\ref{fig:LocalR} provides the structural origin: alongside a near-isostatic carbon--oxygen backbone, UiO-66 contains a strongly over-constrained metal-node population at very negative $\nu_i$, highlighting pronounced heterogeneity in local rigidity.

To confirm the geometric origin of the UiO-66 zero modes, we systematically introduce auxiliary long-range bonds by tuning the cutoff parameter $\tau$ (Fig.~\hyperref[fig:Ui-kicked]{\ref*{fig:Ui-kicked}}). As $\tau$ decreases, the constraint density increases; assigning finite stiffness to these auxiliary bonds lifts the zero modes to finite frequencies, producing soft collective branches that progressively fill the low-frequency spectrum.

Taken together, these case studies show that global constraint counting alone cannot predict rigidity in porous crystalline materials. ABIXOZ behaves as a generically rigid network, whereas IKEBUV01 and UiO-66 derive their mechanical character from accidental modes and constraint dependencies that are only apparent when global indices are complemented by spectral (IPR) and local ($\nu_i$) diagnostics.

\section{Discussion}

Our results reveal that many MOFs occupy a mechanically marginal regime in a constraint-based description. Although the majority of frameworks are formally over-constrained at, their node and linker geometries can introduce \textit{accidental geometric modes} that make subsets of constraints effectively dependent. The resulting excess zero modes are not necessarily distributed uniformly throughout the framework; instead, Figs.~\ref{fig:Phonons} and \ref{fig:LocalR} show that marginality is often highly heterogeneous and can be concentrated in specific substructures. In this sense, global Maxwell--Calladine counts are informative but incomplete: frameworks with similar global $\nu$ can exhibit qualitatively different low-frequency dynamics depending on where constraint redundancy occurs and which atoms participate in the soft subspace. Only a smaller subset of materials, exemplified by ABIXOZ, exhibits fully independent constraints and behaves as a genuinely rigid network for which global counting reliably predicts mechanical response.

The representative cases further show that the \emph{character} of the zero-mode subspace matters as much as its \emph{size}. In IKEBUV01, the zero-like motion is overwhelmingly dominated by hydrogen, indicating that the soft subspace is largely confined to linker-associated, peripheral degrees of freedom. By contrast, UiO-66 exhibits a large number of accidental zero modes while remaining dominated by hydrogen but with a substantial carbon component and only minor oxygen participation, consistent with soft motion that extends further into linker environments even while coupling weakly to the metal--oxygen node. This distinction helps clarify why frameworks that are similarly marginal by global indices can nonetheless differ in how mechanically consequential their soft modes are, depending on whether the zero-mode subspace remains terminal/peripheral or involves heavier framework atoms.

UiO-66, which is widely regarded as a mechanically stable MOF~\cite{UiO-66-stable,UiO-66-stable2}, it is not simply rigid. To understand how the zero modes we identify in our idealized model can produce stability when going beyond our idealizations, we have to understand the limitations of our approach.

While the present rigidity-based analysis provides these mechanistic insights, it is subject to several approximations. The harmonic model captures only small-amplitude motions, and quantitative accuracy depends on the UFF4MOF parameterization. The omission of nonbonded forces and entropic contributions limits the description of large deformations and finite-temperature effects. Despite these simplifying assumptions, the correspondence with first-principles calculations for benchmark systems indicates that the model captures the essential degree-of-freedom structure that governs phonon modes. 

Taken together, these results bridge the gap between simple topological metrics and fully atomistic simulations. Maxwell-based counting alone cannot predict mechanical stability in crystalline porous materials. Rigidity matrix analysis, however, determines whether geometric constraints are independent or redundant, allowing rapid identification of both genuinely rigid and potentially fragile structures. This capability establishes a scalable and interpretable foundation for mechanical screening across large materials databases.

Finally, the prevalence of near-isostatic frameworks revealed here suggests a broader connection between MOF mechanics and ideas from topological mechanics. The parallels to isostatic lattices explored by Kane and Lubensky~\cite{Kane2013} raise the possibility that related topological features, including protected boundary modes, may be realizable in complex porous crystals.

\section*{Acknowledgments}
This material is based upon work supported by the National Science Foundation under Grant No. DMR-1940243.

\appendix

\section{Fourier Transformed Rigidity Matrix}
\label{Appendix:FT}

To analyze vibrational modes in periodic structures, we work in reciprocal
space by Fourier transforming the atomic displacements. Each atom is indexed
by its unit cell position $\mathbf{R}_m$ and an internal label $\alpha$
identifying the atom within the cell. The displacement of atom $(m,\alpha)$
is expanded as
\begin{equation}
    \mathbf{u}_{m\alpha}
    = \frac{1}{\sqrt{N}} \sum_{\mathbf{k}}
      \mathbf{u}_{\alpha}(\mathbf{k})\,
      e^{i \mathbf{k} \cdot \mathbf{R}_m},
    \label{eq:FT_u}
\end{equation}
where $N$ is the number of unit cells, $\mathbf{k}$ runs
over wave vectors in the first Brillouin zone, and
$\mathbf{u}_{\alpha}(\mathbf{k})$ are the Fourier components of the
displacements.

The bond extension for a linear constraint (spring) connecting atom
$(m,\alpha)$ to atom $(n,\beta)$ is
\begin{equation}
    e_{\ell}
    = \hat{\mathbf{r}}_{m\alpha,n\beta} \cdot
      \left( \mathbf{u}_{m\alpha} - \mathbf{u}_{n\beta} \right),
\end{equation}
where $\hat{\mathbf{r}}_{m\alpha,n\beta}$ is the unit vector along the bond.
Substituting the Fourier representation gives
\begin{align}
    e_{\ell}
    &= \hat{\mathbf{r}}_{m\alpha,n\beta} \cdot
       \frac{1}{\sqrt{N}} \sum_{\mathbf{k}}
       \left[ \mathbf{u}_{\alpha}(\mathbf{k}) e^{i \mathbf{k} \cdot \mathbf{R}_m}
       - \mathbf{u}_{\beta}(\mathbf{k}) e^{i \mathbf{k} \cdot \mathbf{R}_n} \right] \nonumber \\
    &= \frac{1}{\sqrt{N}} \sum_{\mathbf{k}}
       e^{i \mathbf{k} \cdot \mathbf{R}_m}\,
       \hat{\mathbf{r}}_{m\alpha,n\beta} \cdot
       \left[ \mathbf{u}_{\alpha}(\mathbf{k})
       - \mathbf{u}_{\beta}(\mathbf{k}) e^{i \mathbf{k} \cdot \boldsymbol{\delta}_{\ell}} \right],
\end{align}
where $\boldsymbol{\delta}_{\ell} = \mathbf{R}_n - \mathbf{R}_m$ is the
lattice vector difference between the two connected unit cells. Thus the
bond extension at wave vector $\mathbf{k}$ depends only on the displacements
$\mathbf{u}_{\alpha}(\mathbf{k})$ and $\mathbf{u}_{\beta}(\mathbf{k})$ at
the same $\mathbf{k}$.

Collecting all constraint extensions into a vector $\mathbf{e}(\mathbf{k})$
and the corresponding displacements into $\mathbf{u}(\mathbf{k})$, we define
the Fourier transformed rigidity matrix $\mathbf{R}(\mathbf{k})$ by
\begin{equation}
    \mathbf{e}(\mathbf{k}) = \mathbf{R}(\mathbf{k})\,\mathbf{u}(\mathbf{k})
    \label{eq:FT_R}
\end{equation}
Substituting this into the potential energy expression,
\begin{equation}
    V = \tfrac{1}{2}\,\mathbf{u}^{\mathrm{T}} (R^{\mathrm{T}} K R) \mathbf{u},
\end{equation}
and using the expansion \eqref{eq:FT_u}, we obtain
\begin{equation}
    V = \frac{1}{2} \sum_{\mathbf{k}}
        \mathbf{u}^\dagger(\mathbf{k})\,
        \mathbf{R}^\dagger(\mathbf{k})\,K\,
        \mathbf{R}(\mathbf{k})\,\mathbf{u}(\mathbf{k}),
\end{equation}
where $K$ is the diagonal matrix of constraint spring constants and
$\dagger$ denotes Hermitian conjugation.

This motivates the definition of the mass weighted dynamical matrix,
\begin{equation}
    \mathbf{D}(\mathbf{k}) =
    M^{-1/2}\,\mathbf{R}^\dagger(\mathbf{k})\,K\,
    \mathbf{R}(\mathbf{k})\,M^{-1/2},
\end{equation}
where $M$ is the diagonal mass matrix. The phonon modes follow from the
eigenvalue equation
\begin{equation}
    \mathbf{D}(\mathbf{k})\,\mathbf{u}(\mathbf{k})
    = \omega^2(\mathbf{k})\,\mathbf{u}(\mathbf{k}),
\end{equation}
with $\omega^2(\mathbf{k})$ the eigenvalues corresponding to the squared
angular frequencies of the phonon modes.

\section{Code and Data Availability}
\label{Appendix:Data}
All computational scripts and input files used in this work are available at:

\begin{center}
\url{https://github.com/coentropy/MOFs}
\end{center}

Structures with CIF sizes above 35 kB were excluded to avoid memory bottlenecks. Neighbor environments were identified using \texttt{CrystalNN}\cite{crystalNN} with a
search cutoff of 25~\AA. For the main dataset we used the default, chemically
aligned settings with \texttt{x\_diff\_weight} = 1.5, which produce coordination
numbers consistent with known metal and linker environments. For the UiO-66
tuning in Fig.~\ref{fig:Ui-kicked}, we instead reran \texttt{CrystalNN} with
chemical heuristics effectively turned off (\texttt{distance\_cutoffs} = None,
\texttt{weighted\_cn} = True, \texttt{x\_diff\_weight} = 0) and then retained
only neighbors with $\mathrm{weight} \ge \tau$. In this case the network is
controlled purely by geometry through the $\tau$ filter: bonds and angles present
in both the default and $\tau$ filtered networks are treated as primary UFF4MOF
constraints, while those that appear only in the $\tau$ filtered network are
interpreted as auxiliary long range constraints and assigned uniform spring
constants (see the repository for specific values). The Maxwell–Calladine index was consistently evaluated at the $\Gamma$-point, while phonon dispersions were sampled along Setyawan--Curtarolo~\cite{Setyawan_2010} high-symmetry paths using 100 evenly spaced points per segment.

To validate the consistency between the UFF4MOF equilibrium bond distances and the geometric bond lengths extracted from the MOF structures, we compared the summed UFF radii $r_{ij}$ against the actual interatomic separations obtained from the crystallographic coordinates. (Note: in the UFF expression $r_{ij} = r_i + r_j + r_{bo} + r_{en}$, $r_{bo}$ and $r_{en}$ are the bond–order and electronegativity corrections; here we set them to zero since their contributions are small for the bonds considered.) Across all linear bonds in the dataset, we found that the UFF equilibrium distances are systematically larger than the geometric bond lengths by approximately 5–7\% on average. This suggests that the UFF radii modestly overestimate the bond lengths present in the structures, but remain within a relatively narrow range, so we adopt them as rest lengths in the rigidity model for the present, approximate treatment. \\

\section{Silicon Benchmark}
\label{Appendix:Phonons}

\begin{figure}[h!]
\centering
\begin{tikzpicture}
\node[inner sep=0] (img) {\includegraphics[width=\linewidth,trim=4 6 4 2,clip]{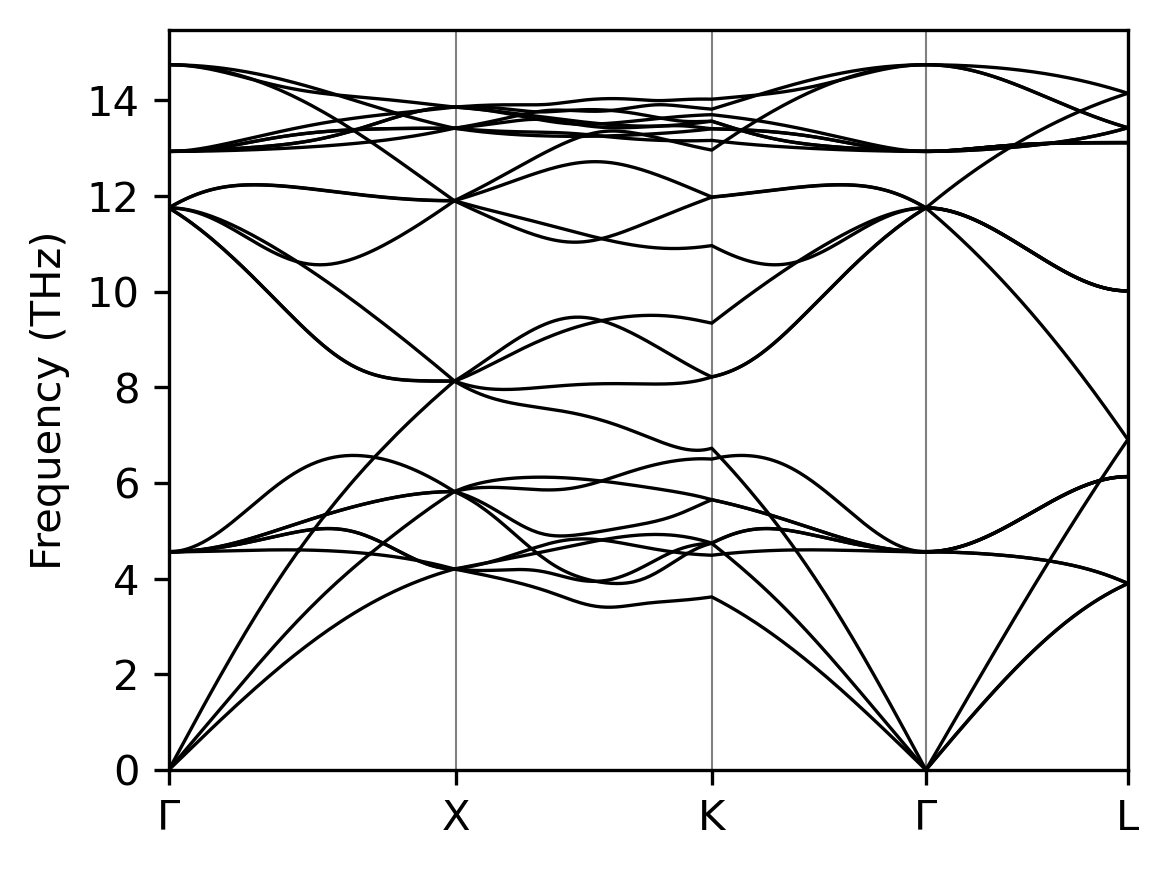}};
\node[anchor=north west, fill=white, inner sep=2pt] at ([xshift=+5pt,yshift=2pt]img.north west) {\bfseries (A)};
\end{tikzpicture}
\vspace{0.4em} 

\begin{tikzpicture}
\node[inner sep=0] (img) {\includegraphics[width=\linewidth,trim=4 6 4 2,clip]{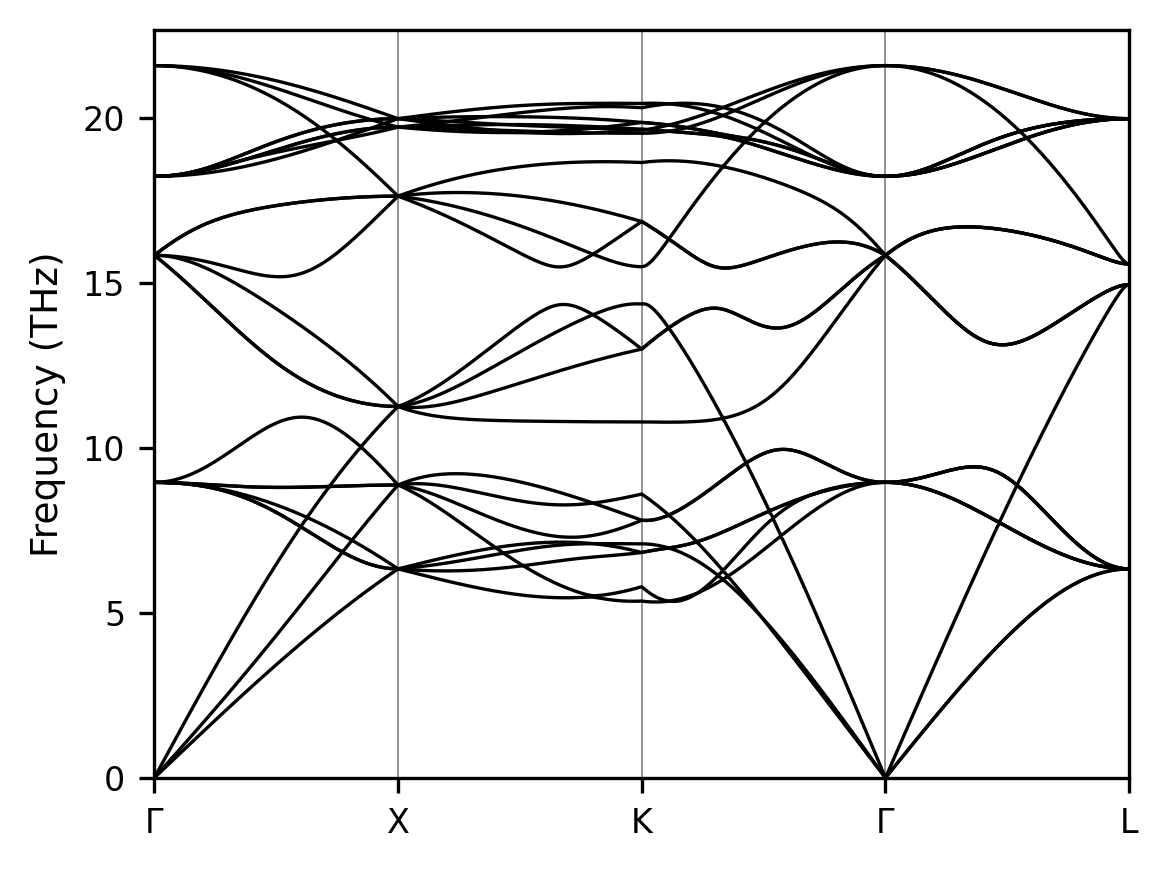}};
\node[anchor=north west, fill=white, inner sep=2pt]
at ([xshift=5pt,yshift=10pt]img.north west) {\bfseries (B)};
\end{tikzpicture}
\caption{(\textbf{A}) Phonon dispersion of a silicon conventional cell calculated by DFT in Quantum Espresso, and (\textbf{B}) the corresponding result from our rigidity-based model. The high-symmetry points coincide, and the overall branch shape is reproduced accurately. Frequencies from the rigidity model are slightly higher, reflecting its stiffer effective force constants relative to the ab initio calculation.}
\label{fig:Silicon}
\end{figure}

Since our present implementation builds the rigidity matrix from an explicit real-space reference cell containing a complete set of distinct bond and angle constraints, we compute the phonon dispersion of silicon using the standard eight-atom conventional cubic cell rather than the two-atom primitive cell. The comparison in Fig.~\ref{fig:Silicon} highlights several important features. The overall frequency range predicted by the rigidity model is higher than in the DFT reference calculation, reflecting the stiffer effective force constants in the harmonic spring network. Along the $\Gamma \rightarrow X$ path, however, the dispersions are in close agreement, with comparable band shapes and similar degeneracies at the high-symmetry points. Along the $X \rightarrow K$ path the two spectra begin to diverge, yet the qualitative structure is still preserved. The most noticeable differences appear along the $\Gamma \rightarrow L$ path, where the relative ordering and spacing of branches do not fully align with the DFT results. Despite these discrepancies, the simplified rigidity-based model reproduces the principal features of the silicon phonon dispersion with reasonable accuracy. This agreement supports the use of the approach for capturing the dominant degree of freedom structure in more complex materials.

\bibliography{references}

\end{document}